\documentclass[10pt,aps,prb,twocolumn,superscriptaddress,floatfix,showpacs,longbibliography]{revtex4-1}

\usepackage{graphicx}
\usepackage{dcolumn}
\usepackage{bm}
\usepackage{hyperref}

\usepackage[usenames,dvipsnames,table]{xcolor}
\usepackage[version=3]{mhchem} 
\usepackage{braket}
\usepackage{multirow}
\usepackage{soul}

\definecolor{amaranth}{rgb}{0.9, 0.17, 0.31}
\definecolor{battleshipgrey}{rgb}{0.52, 0.52, 0.51}
\definecolor{pastelred}{rgb}{1.0, 0.41, 0.38}
\definecolor{royalpurple}{rgb}{0.47, 0.32, 0.66}
\definecolor{mark}{rgb}{0.85, 0.9, 1}
\definecolor{tealgreen}{HTML}{339961}

\newcommand{\state}[1]{\vcenter{\hbox{\includegraphics[totalheight=0.037\columnwidth]{off-site_state_#1.png}}}}
\newcommand{\eqstate}[1]{\vcenter{\hbox{\includegraphics[totalheight=0.05\columnwidth]{off-site_state_#1.png}}}}

\hypersetup{hidelinks}
\sethlcolor{mark}

\begin{document}
	
\title{Quantum Nanoskyrmions}

\author{O. M. Sotnikov}
\affiliation{Theoretical Physics and Applied Mathematics Department, Ural Federal University, Mira Str. 19, 620002 Ekaterinburg, Russia}

\author{V. V. Mazurenko}
\email{vmazurenko2011@gmail.com}
\affiliation{Theoretical Physics and Applied Mathematics Department, Ural Federal University, Mira Str. 19, 620002 Ekaterinburg, Russia}

\author{J. Colbois}
\affiliation{Institute of Physics, \'Ecole Polytechnique F\'ed\'erale de Lausanne (EPFL), CH-1015 Lausanne, Switzerland}

\author{F. Mila}
\affiliation{Institute of Physics, \'Ecole Polytechnique F\'ed\'erale de Lausanne (EPFL), CH-1015 Lausanne, Switzerland}

\author{M. I. Katsnelson}
\affiliation{Radboud University, Institute for Molecules and Materials, Nijmegen, Netherlands}
\affiliation{Theoretical Physics and Applied Mathematics Department, Ural Federal University, Mira Str. 19, 620002 Ekaterinburg, Russia}

\author{E. A. Stepanov}	
\affiliation{Radboud University, Institute for Molecules and Materials, Nijmegen, Netherlands}
\affiliation{Theoretical Physics and Applied Mathematics Department, Ural Federal University, Mira Str. 19, 620002 Ekaterinburg, Russia}

\date{\today}
	
\begin{abstract}
Skyrmions in condensed matter physics appear as {\it classical} topological spin structures. This nontrivial state can be obtained solving the corresponding micromagnetic model, where the magnetization is treated as a continuous classical vector field.  Here, we introduce a new concept of {\it quantum} skyrmions of a nanoscale size. This distinct magnetic state can be formed in spin-$1/2$ low dimensional magnets characterized by a strong Dzyaloshinskii-Moriya interaction. This concept can also be applied to usual spin systems characterized by a nonzero exchange interaction if the quantum solution of the problem is accessible. 
To perform a complete characterization of such a quantum nanoskyrmion state, we analyze basis functions giving the largest contribution to the ground state at different values of external magnetic field and temperature. We observe that the quantum skyrmionic state is stabilized even when the corresponding classical skyrmionic solution has already undergone a phase transition towards the polarized ferromagnetic state.
\end{abstract}

\maketitle

The evident progress in the development of experimental techniques for the observation of magnetic skyrmions, topologically protected spin structures~\cite{bogdanov1989thermodynamically},
% BOGDANOV1994255
poses new challenges for theory and numerical simulations of ordered magnetic phases~\cite{ Muhlbauer915, PhysRevLett.102.186602, PhysRevB.81.041203, yu2010real, nagaosa2013topological,TsirlinCu}. Nowadays, skyrmions are mostly discussed in the context of spintronics, where these stable magnetic structures are proposed as bits in magnetic memory devices~\cite{Jonietz1648, fert2013skyrmions, Tomasello:2014qd}. The need to store more and more information requires the development of ultradense memories. This motivates the investigation of skyrmions of the nanoscale size. The progress in this direction already made the experimental observation of skyrmions with the characteristic size of a few nanometers~\cite{Wiesendanger, Romming636} possible. Additionally, nanoskyrmions were theoretically predicted in frustrated magnets~\cite{PhysRevLett.108.017206, leonov2015multiply, PhysRevB.93.064430}, narrow band Mott insulators under high-frequency light irradiation~\cite{PhysRevLett.118.157201}, and Heisenberg-exchange-free systems~\cite{skyrm_classic}. On such small characteristic length scale compared to the lattice constant, quantum effects cannot be neglected. In this respect, the numerical study of classical spin models can no longer be considered as an exhaustive solution of the problem. This also concerns low-dimensional systems with small spin (e.g. $S=1/2$) and itinerant magnetic systems with not very well localized magnetic moment, where quantum fluctuations play a crucial role.

The problem of the quantum description of skyrmions is rooted in the fact that in most of the cases these topological structures emerge as the result of magnetic frustration or competing magnetic interactions. This restricts the applicability of quantum Monte Carlo methods due to the sign problem~\cite{sign_problem}. On the other hand, a quantum skyrmionic problem necessarily requires a cluster solution, since the local magnetization of a skyrmion points in different directions at different lattice sites. Exact diagonalization (ED) based methods only allow for an accurate quantum description of small clusters. Standard approaches are limited to sizes around 30 sites in the case of spin-$1/2$ systems. Advanced ED techniques allow to reach 50 sites~\cite{Andreas,Andreas1} but require supercomputers. This is at least two orders of magnitude smaller than that required for the description of skyrmions in transition metal compounds.

The introduction of a quantum analog of the classical skyrmionic structure is also complicated by the fact that it is totally unclear how the quantum solution of the problem can be mapped onto a classical object in this particular case. This issue can simply be explained by the following example. It is widely known that the ground state of the classical Heisenberg model on a triangular plaquette with the antiferromagnetic (AFM) exchange interaction between neighboring sites is given by the so-called 120$^{\circ}$ spin state. However, the latter cannot be restored from the solution of the quantum Heisenberg model. In this case the quantum ground state is degenerate due to the spin frustration. The same problem occurs when introducing the quantum skyrmion, because the existence of classical skyrmionic structures almost entirely relies on the presence of the frustration due to the Dzyaloshinskii-Moriya interaction (DMI) or other competing magnetic interactions. 

One possible way to approach this problem is to force the quantum problem to behave as a classical one. In the theory of antiferromagnetism, this can be performed by applying a staggered field that selects the classical N\'eel state from the quantum solution of the problem. The same can be achieved with the use of the analog of the quantum Zeno effect~\cite{doi:10.1063/1.523304, joos2013decoherence} by performing repeated local measurements~\cite{PhysRevB.98.014416}. Skyrmions are usually identified by the winding number, which implies the calculation of a three-particle observable. Hence, one has to continuously measure the local three-spin correlation function in order to make the quantum skyrmion behave like a classical object. Unfortunately, this can hardly be done in experiments. 

On the other hand, the description of the skyrmionic problem can be done semiclassically assuming that the magnetization dynamics is dominated by classical magnetic excitations that emerge on top of the symmetry-broken ground state of the system (for details see Ref.~\onlinecite{2018arXiv180702203O} and references therein). Then, the winding number can be calculated using the triangulation of the spin lattice using the approach proposed by Berg and L\"uscher in Ref.~\onlinecite{luscher}. The latter is based on the calculation of the solid angles subtended by three neighbouring spins. Unfortunately, this idea cannot be successfully used in the purely quantum case due the to uncertainty principle, because for the calculation of the solid angle one has to know all three projections of spins simultaneously. 

As a result, quantum effects in magnetic skyrmion models are normally taken into account by means of the Holstein-Primakoff transformation, which only allows to compute quantum corrections to the classical solution. This already led to interesting results reported in Ref.~\onlinecite{PhysRevB.92.245436}. There, the authors demonstrated that quantum spin fluctuations of a noncollinear spin texture produce effective Casimir-like magnetic fields. However, none of the above methods can provide a complete quantum description of the specified problem. 

Here, we report on the first attempt to introduce the concept of {\it purely quantum} skyrmions based on the exact numerical solution of the quantum problem. The idea follows from the example of the quantum ground state of the triangular plaquette given above. Although the AFM ordering on the triangular plaquette cannot be identified directly in the quantum ground state, it is nicely captured by the static spin structure factor whose intensity peaks exactly at the wave vector corresponding to the AFM ordering. Moreover, this holds not only for the symmetry-broken case, but in the paramagnetic phase as well. Here, the AFM intensity appears close to the magnetic instability at a finite energy proportional to the inverse of the order parameter~\cite{Stepanov2018}. Since the classical skyrmion can be seen as the superposition of enclosed spin spirals, the structure factor should reveal intensities at the momenta that are related to the period of these spirals. Therefore, the quantum skyrmion can be identified as {\it a multiple-$q$ state of a quantum system with a special distribution of intensities in the spin structure factor comparable to the classical skyrmionic case.}

\begin{figure}[t!]
	\includegraphics[width=\columnwidth]{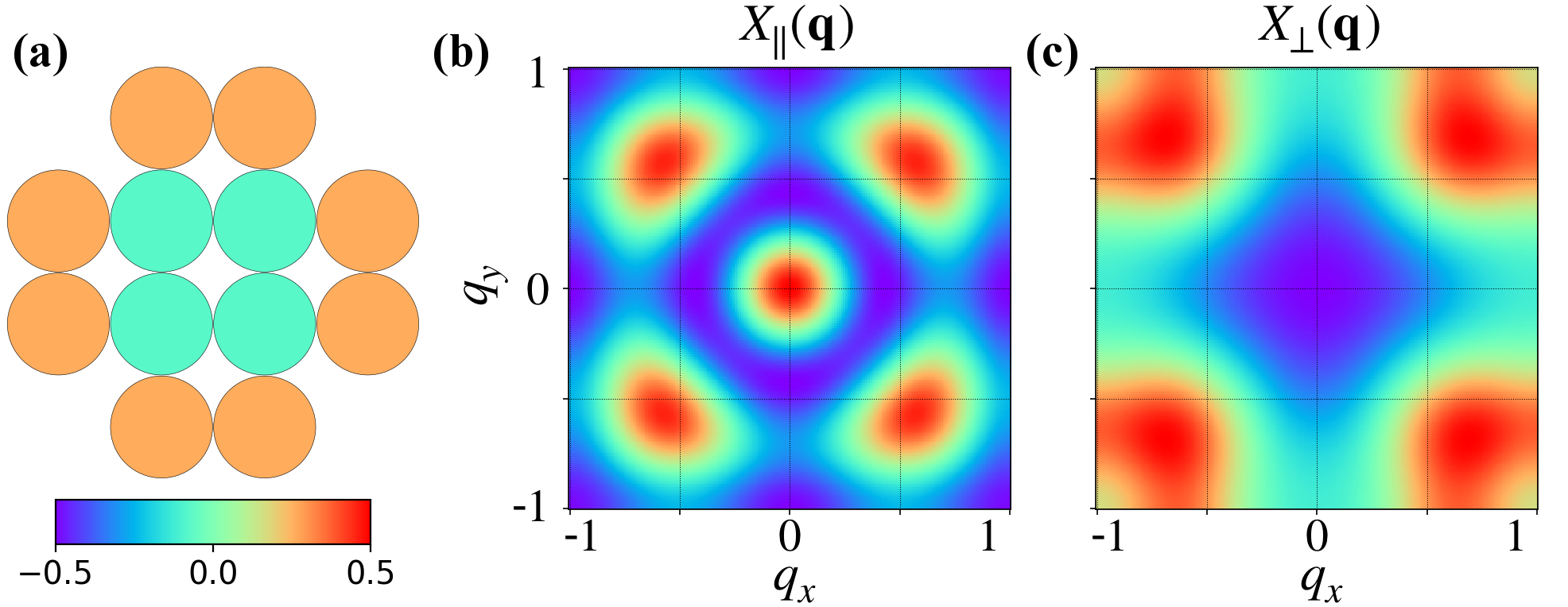}
	\caption {\label{fig:off-site_energy_spectrum} (a) $z$-components of the local magnetization of the off-site cluster obtained from ED for $B = 0.25$ at zero temperature. (b) $X_{\|}(\mathbf{q})$ and (c) $X_{\bot}(\mathbf{q})$ static spin structure factor corresponding to this solution.}
\end{figure}

The exact solution of the quantum problem can be obtained only for a limited number of lattice sites as discussed above. Here, we explore the quantum skyrmionic problem using the quantum Heisenberg-exchange-free Hamiltonian introduced in Ref.~\onlinecite{skyrm_classic}
\begin{equation}
\label{eq:ham}
%\hat{\mathcal{H}}
\hat H = \sum_{i<j} \mathbf{D}_{ij}[\hat {\mathbf{S}}_i\times\hat{\mathbf{S}}_j] - \mathbf{B}\sum_i \hat{\mathbf{S}}_{i},
\end{equation}
where the vector $\mathbf{D}_{ij}$ lies perpendicularly to the bond between sites $i$ and $j$, and describes the Dzyaloshinskii-Moriya interaction. $\mathbf{B}$ is the external magnetic field that is assumed to be along $z$. Such a purely anisotropic form of the spin Hamiltonian leads to very compact topological excitations at the classical level. For instance, the off-site classical nanoskyrmion can be stabilized on a plaquette of 12 sites visualized in Fig.~\ref{fig:off-site_energy_spectrum}\,(a). Therefore, the corresponding spin-$1/2$ quantum problem~\eqref{eq:ham} can be solved numerically exactly by means of the full diagonalization approach~\cite{ALPS1}. We show that the quantum skyrmion state can be fully described by analyzing the components of the eigenstates of Eq.~\ref{eq:ham}. We identify the basis multi-spin functions of the quantum Hamiltonian that can be associated with the classical nanoskyrmion species obtained in Ref.~\onlinecite{skyrm_classic}. Note that the rather exotic Heisenberg-exchange free model is considered here only for the sake of the exact quantum solution of the problem. The full quantum skyrmionic solution can also be obtained for a conventional spin Hamiltonian with nonzero exchange interaction. However, it will imply numerical simulations on much larger spin clusters.

{\it Quantum nanoskyrmion} --- 
We start with the main result of our study, namely the demonstration that a quantum nanoskyrmion state is stabilized on the 12-site plaquette (Fig.~\ref{fig:off-site_energy_spectrum}). Such a quantum nanoskyrmion appears as the ground state of the quantum DMI model Hamiltonian~\eqref{eq:ham} at finite magnetic field. For instance, results shown in Fig.~\ref{fig:off-site_energy_spectrum} are obtained for a magnetic field $B = 0.25\,|\mathbf{D}|$. Hereinafter, the energy is given in units of DMI ($|\mathbf{D}| = 1$). As discussed in the introduction, the winding number cannot be calculated due to the uncertainty in the determination of the spin components. For this reason, we use the calculated longitudinal $X_{\parallel}$ and transverse $X_{\perp}$ spin structure factor in order to detect the skyrmionic state. They are given by the following expression
\begin{eqnarray}
X_{\| (\bot)} (\mathbf{q}) = \frac{1}{N}\sum_{ij} X_{\| (\bot)}(\mathbf{R}_{ij}) e^{-i \mathbf{q} \cdot \mathbf{R}_{ij}},
\end{eqnarray}
where $N$ denotes the number of spins, and the real-space spin-spin correlation functions are defined as
\begin{align}
X_{\|}(\mathbf{R}_{ij}) &= \frac{1}{Z} \sum_{n} \langle n|\hat S_{i}^z \hat S_{j}^z|n \rangle e^{-\beta E_{n}},\\
X_{\bot}(\mathbf{R}_{ij}) &= \frac{1}{Z} \sum_{n} \langle n|\hat S_{i}^x \hat S_{j}^x + \hat S_{i}^y \hat S_{j}^y|n \rangle e^{-\beta E_{n}}, \nonumber
\end{align}
where $Z$, $n$, and $E_{n}$ are the partition function, eigenfunction, and eigenvalue of the DMI Hamiltonian~\eqref{eq:ham}, respectively.

\begin{figure}[!t]
	\includegraphics[width=1\linewidth]{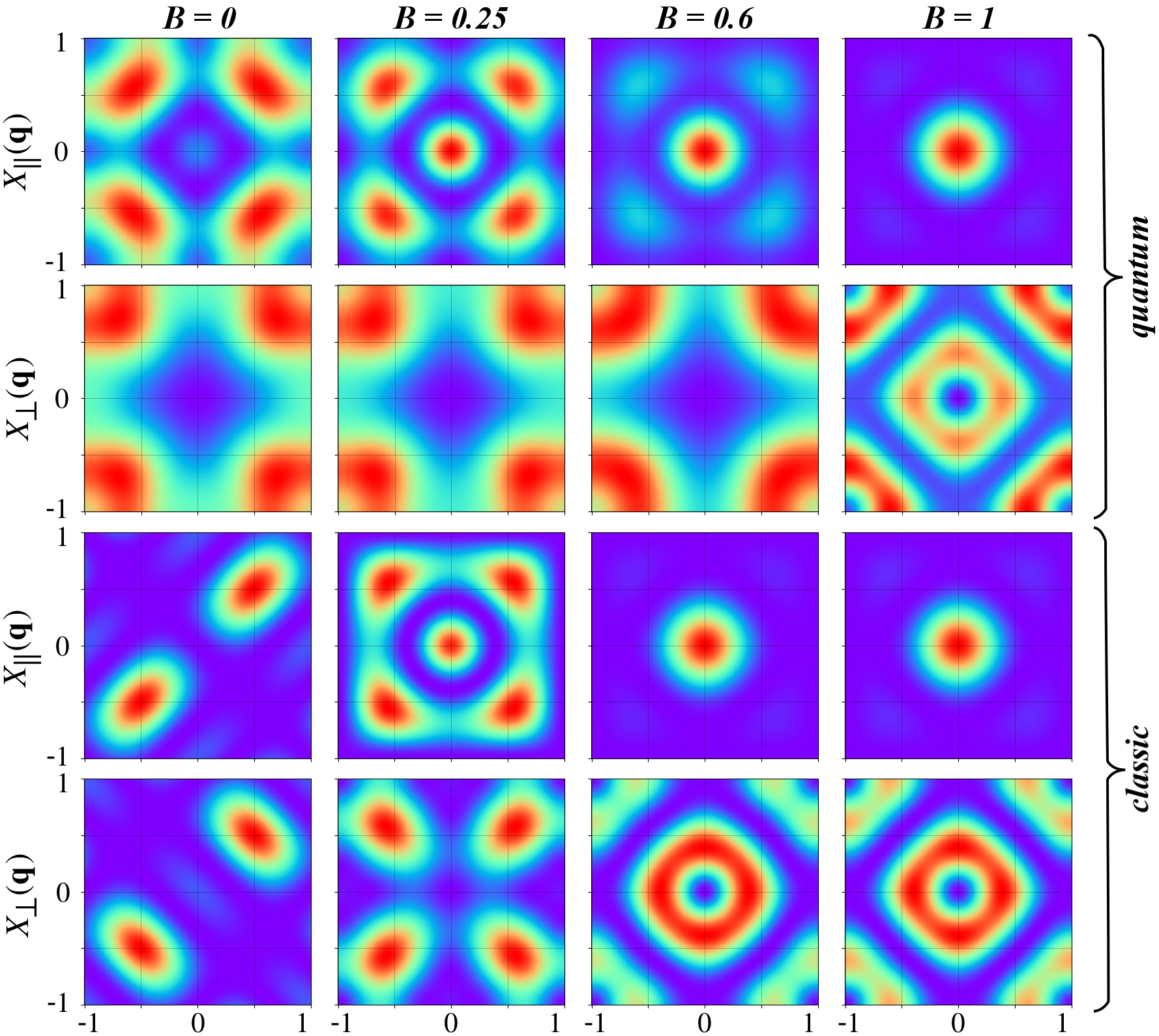}
	\caption {\label{comp_suscep} (top) Quantum and (bottom) classical spin structure factor calculated at different magnetic fields $B$ at zero temperature. Values of the magnetic field are given in units of DMI ($|\mathbf{D}| = 1$).}
\end{figure}

From Fig.~\ref{fig:off-site_energy_spectrum} it can be seen that the quantum system is characterized by a multispiral (multiple-$q$) state. Here, maximal intensities are located at ${\bf q} = 0$ and ${\bf q} = (\pm q, \pm q)$ with $q = 0.56 \pi$. The intensity at ${\bf q} = 0$ reflects the ferromagnetic ordering along the cluster (skyrmionic) boundary. In turn, excitations at ${\bf q} = (\pm q, \pm q)$ correspond to two enclosed spin spirals that appear in the system. Such behaviour of the spin structure factor is a precise signature of the skyrmionic state emerging in the system, as discussed above. Our approach to detect the skyrmionic state is similar to the one that is normally used in experiments. For instance, spin-polarized scanning tunneling measurements~\cite{Wiesendanger} reveal surface areas with parallel or antiparallel magnetization to the tip magnetization. The Fourier transform of the magnetization texture is then used for the calculation of structure factors demonstrating multiple-$q$ intensities.  

Importantly, for all values of the magnetic field, the ground state energy of the quantum solution is much lower than that of the classical one. The latter is obtained by minimizing the classical spin Hamiltonian with respect to spin orientations. The corresponding energy difference (for the entire system) obtained for the magnetic field $B=0.25$ is equal to $\Delta E = E_{{\rm qt}} - E_{{\rm cl}}  = - 0.98$. This energy difference is related to quantum fluctuations and corresponds to the zero-point energy in the framework of the quantum spin-wave theory~\cite{PhysRevB.92.245436}. The same spin-wave study has demonstrated another important consequence of quantum fluctuations, namely the partial compensation of the magnetic field effect and the extension of the skyrmionic area in the phase diagram. As we show below, the exact quantum solution reveals an even more intriguing behaviour of the system showing the {\it coexistence} of ferromagnetic and skyrmionic states at magnetic fields larger than the critical one estimated from classical simulations. Remarkably, these two states are present in the system even at magnetic fields approaching twice the critical value.   

The complete evolution of the skyrmionic state on the 12-site plaquette at different values of the magnetic field is visualized in Fig.~\ref{comp_suscep}. At zero magnetic field and temperature, the position of intensities corresponds to a superposition of two spin spirals. As we show below, for each basis state contributing to the ground state, there exists another one with opposite magnetization. At the classical level the system chooses only one single-spiral configuration because the superposition of states is not allowed for the classical solution of the problem. The quantum solution obtained at $B = 0.6$ (Fig.~\ref{comp_suscep}) is characterized by non-zero intensities at $\mathbf{q} = (\pm q, \pm q) $, which is an indication of the skyrmionic state formed in the system. By contrast, the classical solution obtained at the same magnetic field shows a purely ferromagnetic state with only non-zero intensity at ${\bf q}=0$.

\begin{figure}[t!]
	\includegraphics[width=1\linewidth]{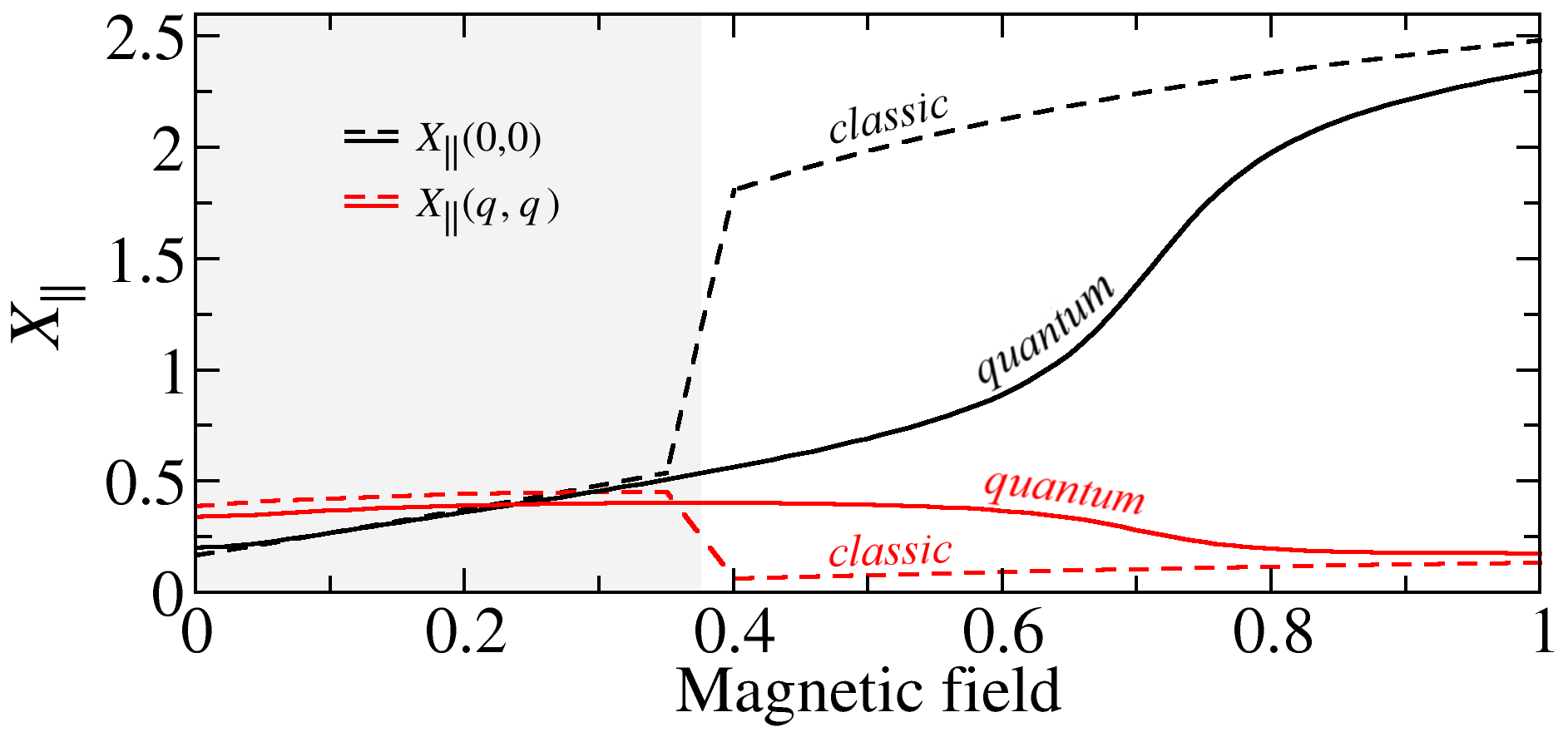}
	\caption{\label{fig:st_magn} Magnetic field dependence of $X_{\|}(0,0)$ (black curves) and  $X_{\|}(q, q)$ (red curves) calculated at zero temperature for the quantum (solid curves) and the classical (dashed curves) problem. The highlighted region indicates the values of the magnetic field corresponding to the classical skyrmionic phase.}
\end{figure}

To give a quantitative description of the transition between skyrmionic and ferromagnetic phases we obtain the values of the maximal intensities of the spin structure factor as a function of the magnetic field (Fig.~\ref{fig:st_magn}). A similar comparison has been performed in~\cite{PhysRevB.58.R14733} showing a completely different behaviour of classical and quantum structure factors as a function of the applied magnetic field. Remarkably, in our case, the maximum values of both quantum and classical spin structure factor show almost the same dependence on the magnetic field up to $B=0.4$. However, for larger magnetic fields the situation is completely different. The classical solution of the problem at magnetic fields larger than $B=0.4$ reveals the ferromagnetic state, as shown by comparing the values of $X_{\|} (0, 0)$ and $X_{\|} (\pm q, \pm q)$. The latter becomes much smaller than the first one and approaches zero at magnetic fields above the critical value. By contrast, in the quantum case the spin structure factor $X_{\|}$ at the $(\pm q, \pm q)$ point does not go to zero for all considered values of the magnetic field. Based on this result, we arrive at the very important conclusion that there exists a non-zero probability to find the quantum system in the skyrmionic state even at large magnetic field up to $B\simeq0.7$. Below, we explain this result by analyzing the components of the ground state wavefunction.   

\begin{figure}[t!]
	\includegraphics[width=1\linewidth]{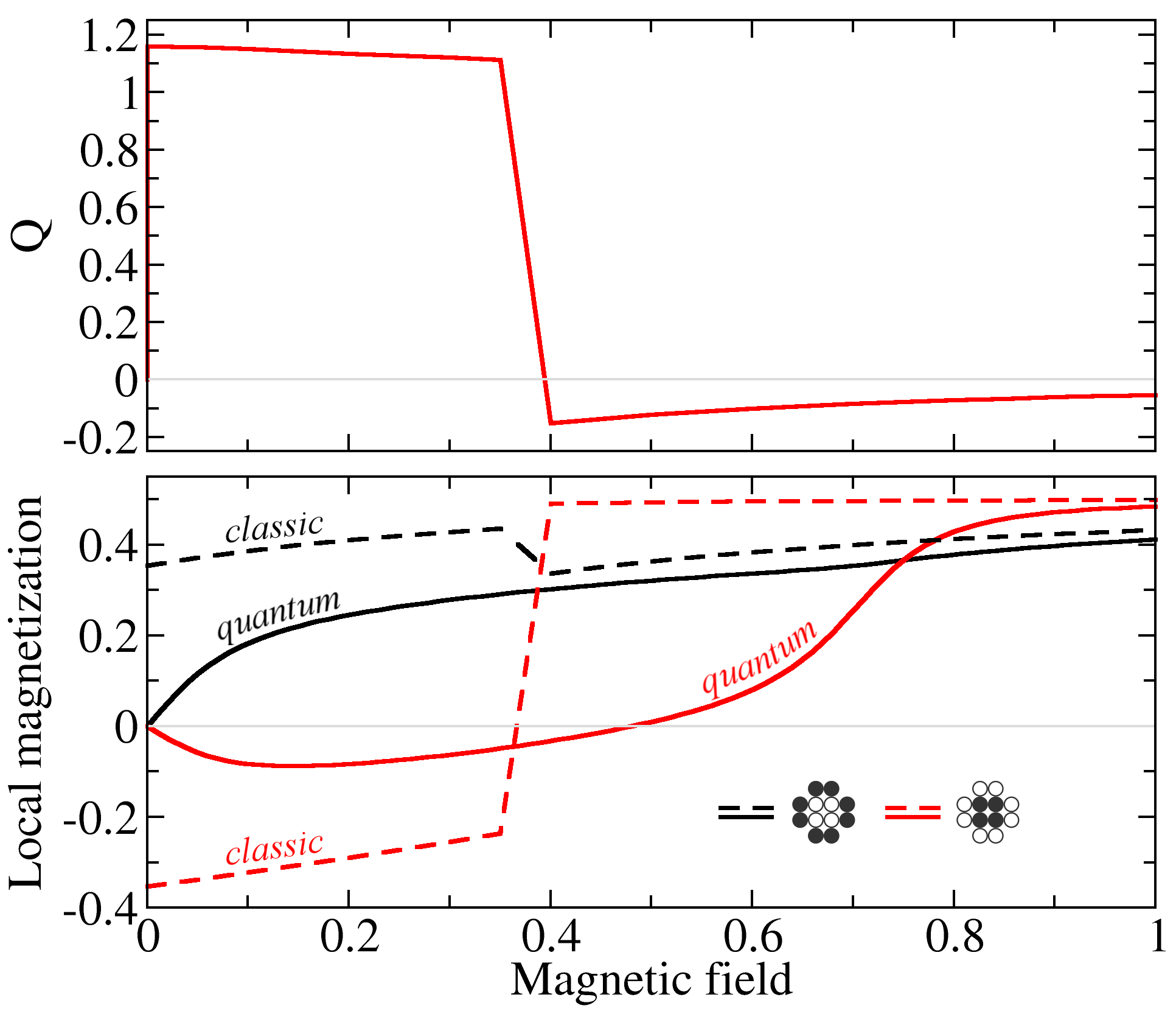}
	\caption{\label{fig:qnum} (top) Skyrmion number calculated for the classical DMI Hamiltonian defined on the 12-site plaquette. (bottom) Magnetization (per site) of central (red lines) and border (black lines) sites as a function of the magnetic field calculated using the minimization of the classical energy (dashed lines) and ED (solid lines) simulations. 
	}
\end{figure}
%\begin{figure}[b!]
%	\includegraphics[width=\columnwidth]{local_magnetization_T_g.png}
%	\caption{\label{fig:lmag} Magnetization (per site) of central (solid lines) and border (dashed lines) sites as a function of the magnetic field calculated at different temperatures. Gray solid line denoting zero values was added for clarity.}
%\end{figure}

The solution of the DMI model with classical spins of length $S=1/2$ by means of the total energy minimization with respect to the spin orientations at zero temperature reveals that the skyrmionic state is stabilized in the range of magnetic fields $0 < B \le 0.4$ (Fig.~\ref{fig:qnum} top). Here, one can see that in the classical case the topological charge correlates with the value of the $z$-component of the central spin (magnetization) of the 12-site cluster. While $\langle \hat{S}^z \rangle$ is negative (the corresponding spin is antiparallel to the magnetic field), the topological charge is nonzero. Previously, a similar approach based on the analysis of the local magnetization was used in Refs.~\onlinecite{Silva, Stamps}. By contrast, in the quantum case the negative magnetization of core spins on the plaquette cannot be used to detect the skyrmionic state in the system. From Fig.~\ref{fig:qnum} (bottom) one can see that the critical magnetic field at which the magnetization of central spins becomes positive is $B\simeq0.5$. At the same time, the spin structure factor demonstrates the multispiral state up to $B\simeq0.7$. Remarkably, the calculated value of the magnetization for the quantum model is strongly suppressed (about 3 times) compared to the classical solution. This is another signature of quantum fluctuations contributing to the skyrmionic solution of this spin-$1/2$ system. A similar reduction of the magnetic moment due to quantum fluctuations was predicted in Ref.~\onlinecite{PhysRevB.92.245436}. However, in our case the effect is more pronounced.  

\begin{figure}[t!]
	\includegraphics[width=\linewidth]{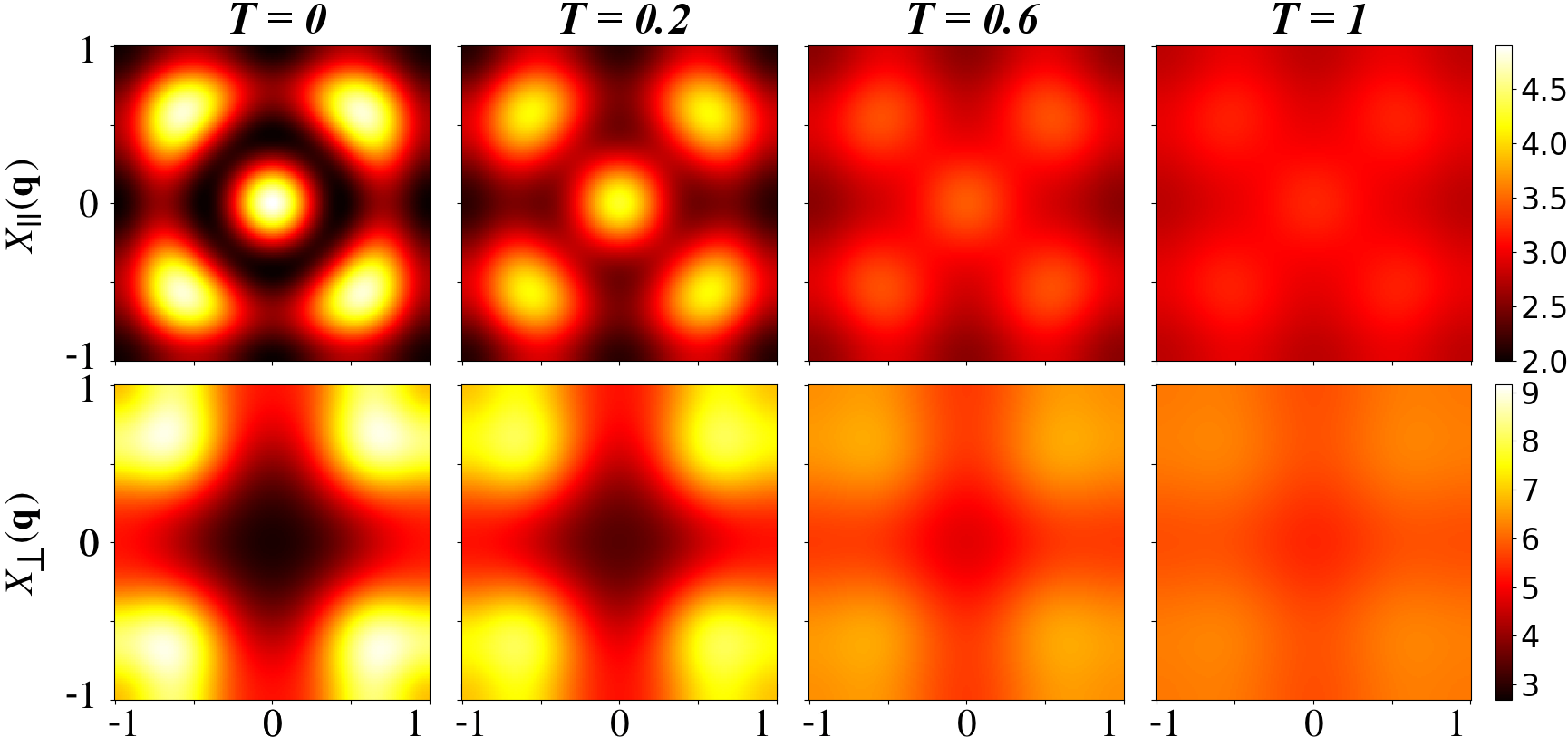}
	\caption{\label{fig:heating} Temperature evolution of the spin structure factor calculated 
		%for quantum DMI model 
		at $B = 0.25$ (skyrmionic regime).}
\end{figure}

In contrast to the previous work~\cite{PhysRevB.92.245436} based on the quantum spin-wave theory focused on zero temperature, we have studied the temperature evolution of the proposed quantum nanoskyrmionic state. Fig.~\ref{fig:heating} shows the structure factors calculated at finite temperatures at $B=0.25$. One can see that the maxima at ${\bf q} = 0$ and ${\bf q} = (\pm q, \pm q)$ can be observed even at $T=1$, which means that the quantum nanoskyrmion is stable against temperature fluctuations. Below we provide a microscopic explanation of these temperature and magnetic field dependencies. For that we analyze the components of the ground state and of the first excited state. 

%\begin{figure}[t!]
%	\includegraphics[width=\columnwidth]{spectra_only.png}
%	\caption{\label{fig:espec} Energy spectrum of off-site cluster. Black and blue solid lines denote two lowest eigenenergies obtained from ED and the classical solution of the DMI Hamiltonian, respectively. The dashed green line indicates the magnetic field value at which the crossing of two lowest levels occurs. Periodic boundary conditions were applied for all calculations.}
%\end{figure}

{\it Analysis of eigenstates} --- 
At zero magnetic field, the largest contributions to the ground and first excited states are 
\begin{align}
		\left|\Psi_0\right|^2  = \alpha \big(&\eqstate{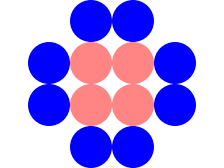} + \eqstate{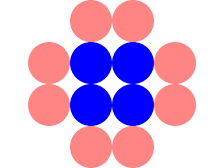}\big) + \beta \big(\eqstate{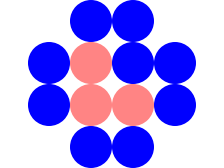} + \eqstate{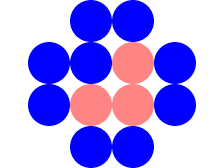}+\eqstate{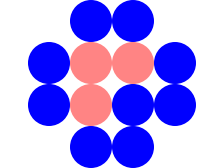}+\eqstate{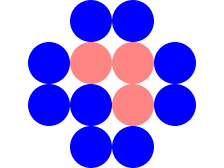} \notag\\ 
           + &\eqstate{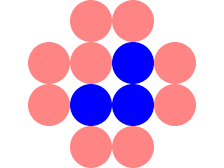} + \eqstate{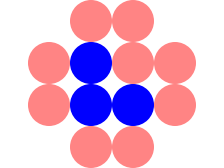}+\eqstate{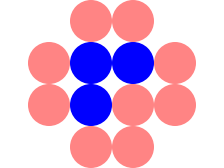}+\eqstate{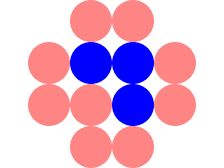}\big) + \ldots \notag\\
    \left|\Psi_1\right|^2 =
    \eta \big(&\eqstate{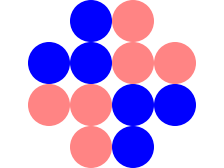} + \eqstate{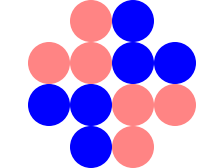}\big) + \xi \big(\eqstate{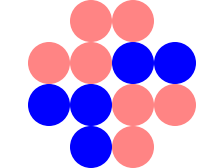} + \eqstate{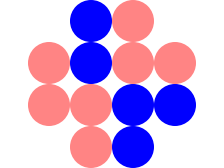} + \eqstate{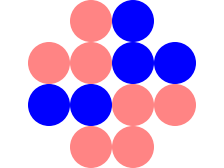} + \eqstate{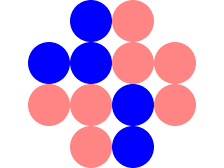} \notag\\
     + &\eqstate{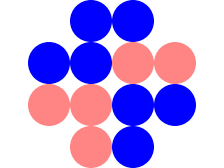} + \eqstate{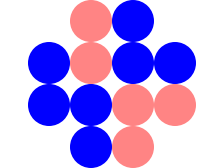} + \eqstate{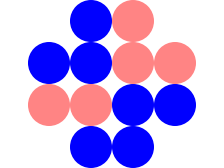} + \eqstate{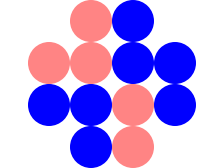}  + \eqstate{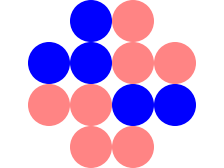} + \eqstate{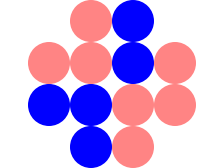} + \eqstate{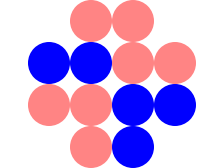} \notag \\
     + &\eqstate{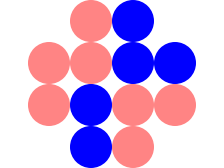} + \eqstate{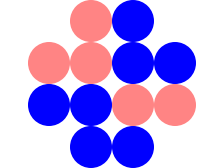} + \eqstate{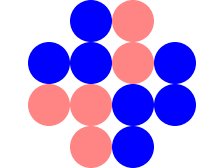} + \eqstate{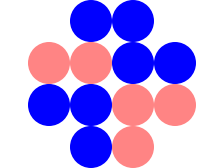} + \eqstate{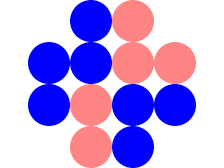}\big) + \ldots,\notag
\end{align}
where probabilities of different basis states are $\alpha=0.0029$, $\beta=0.0016$, $\eta = 0.0027$, and $\xi=0.0018$. The icons denote the square of the corresponding basis functions, $\langle\state{408}  | \state{408} \rangle$ with spin-up (red) and spin-down (blue) sites.

The next step is to analyze the components of the ground state of the quantum system to probe the presence of topological order.
Our main focus is on the 15 basis states $\state{3687}, \state{3943}, \state{3815}, \state{3695},\state{3703}, \state{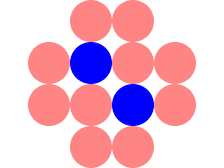}, \state{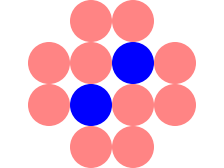}, \state{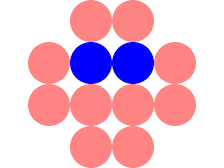}, \state{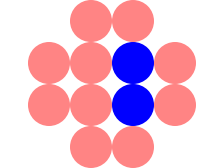}, \state{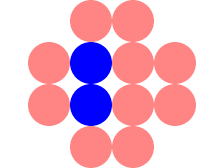}, \state{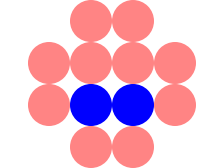}, \state{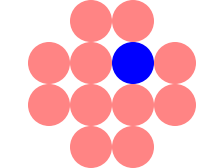}, \state{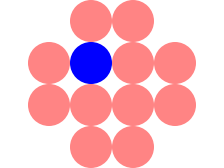}, \state{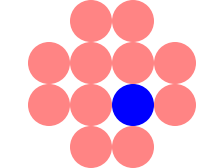}, \state{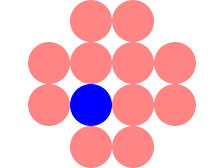}$, where central spins on the plaquette are pointing down, and other spins are pointing up. These states can be divided into 4 groups with one, two, three, and four spins-down in the core of the plaquette, respectively. Basis states within one group are related to each other by a 90$^{\circ}$ rotation around the $z$ axis ($C_{4v}$ symmetry). These quantum spin states of the system are most closely related to classical Monte Carlo solutions reported in Ref.~\onlinecite{skyrm_classic}. For instance, the $\state{3687}, \state{3823}$, and $\state{3967}$ states can be associated with the off-site, bond, and on-site skyrmions, respectively. The contribution of these basis states to the ground state wavefunction can be considered as the main indicator of the quantum nanoskyrmion state stabilized in the plaquette. 

\begin{figure}[t!]
	\includegraphics[width=1\columnwidth]{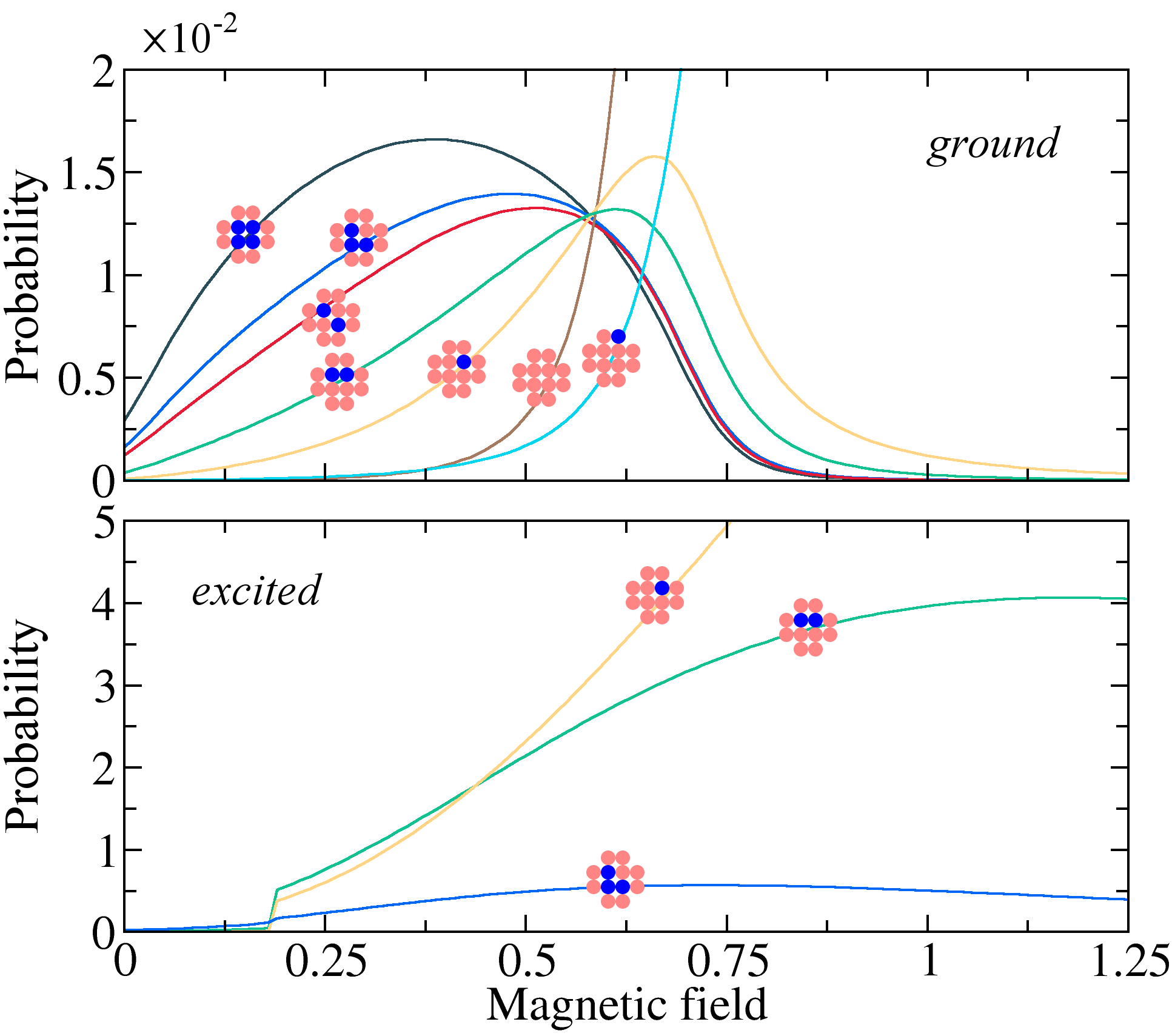}
	\caption{\label{fig:probs} Probabilities of the selected basis functions depending on the magnetic field value for (top) ground and (bottom) first excited states.}
\end{figure}

It follows from Fig.~\ref{fig:probs} (top) that the $\state{3687}$ state gives the largest contribution to the ground state at magnetic field $0<B<0.4$. At the same time, the contribution of the fully polarized $\state{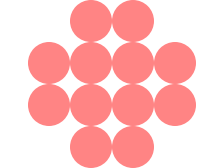}$ state is negligibly small. The latter becomes important for $B > 0.4$. The contribution of the $\state{3967}$ state provides the stability of the skyrmionic state at large fields $B\simeq0.7$. The wavefunction corresponding to the first excited state (Fig.~\ref{fig:probs} bottom) is also characterized by a considerable contribution from skyrmion-like basis states. It follows from Fig.~\ref{fig:probs} that for $B > 0.2$ the $\state{3839}$ and $\state{3711}$ states have the largest probability. At the same time, the fully polarized basis function $\state{4095}$ does not contribute to the first excited state.

The classical solution of the DMI model on small clusters presented in Ref.~\onlinecite{skyrm_classic} has revealed that the off-site, bond, and on-site nanoskyrmions tend to stabilize at different values of the magnetic field. The off-site cluster requires the minimal value of the magnetic field for the stabilization of the skyrmionic phase characterized by a non-zero topological charge. At the same time, the on-site skyrmion is the most compact form. It can be found close to a fully polarized system. We observe a similar trend in the behaviour of basis functions $\state{3687}$ and $\state{3967}$ that we associate with off-site and on-site skyrmions. From Fig.~\ref{fig:probs} (top) one can see that $\state{3687}$ and $\state{3967}$ states give the largest contribution to the ground state at $B = 0.37$ and $B = 0.7$, respectively.

%\chapter{\bf DISCUSSION}

{\it Conclusion} --- 
We have introduced and analyzed a distinct state of a spin system - a quantum nanoskyrmion. We find that the energy of the quantum nanoskyrmion is much smaller than that of its classical analog. Remarkably, we observe that there exists a non-zero probability to find the quantum system in the skyrmionic state even at large magnetic field, where the classical solution of the problem already gives a saturated ferromagnetic solution. We have identified the concrete basis states that are responsible for the formation of the quantum nanoskyrmion. Their relation to classical magnetic configurations is discussed. \\

{\it Acknowledgements ---}
The work of V.V.M., O.M.S., and E.A.S. was supported by the Russian Science Foundation Grant 18-12-00185. The work of M.I.K. was supported by NWO via Spinoza Prize. The work of J.C. and F.M. is supported by the Swiss National Science Foundation. Also, the work was partially supported by the Stichting voor Fundamenteel Onderzoek der Materie (FOM), which is financially supported by the Nederlandse Organisatie voor Wetenschappelijk Onderzoek (NWO). \\

The data that support the findings of this study are available from the corresponding author upon reasonable request.

\end{document}